\documentclass{aa}

\usepackage{natbib}
\usepackage{psfig}

\begin{document}

\title{On the spin evolution of neutron stars in pre-low-mass X-ray binaries}

\titlerunning{Neutron stars in pre-LMXBs}

\author{S.B. Popov
            \inst{1,2}
         }


\institute{Universit\`a di Padova, Dipartimento di Fisica, 
via Marzolo 8, 35131, Padova, Italy
            \and
   Sternberg Astronomical Institute,
Universitetski pr. 13, 119992 Moscow, Russia\\
\email{polar@sai.msu.ru}
}

   \date{}

\abstract{We present a simple model for the evolutionary states
of neutron stars in pre-low-mass X-ray binaries 
during the main sequence stage of donor stars. 
It is shown that for typical parameters
some of the neutron stars in these systems can not accrete matter from winds
of their companions in a stable way. Accretors are found if neutron stars
have magnetic fields about $10^{13}$~G and higher and/or in close systems with
orbital periods about a few days and shorter.
}

\maketitle

\section{Introduction}

The increased ability to observe very faint X-ray sources with modern X-ray
observatories (XMM-Newton, Chandra)
prompted new interest in systems with low accretion rates
$\sim 10^{10}$--$10^{12}$~g~s$^{-1}$ like isolated neutron stars (NSs) or
NSs in detached binaries. Among isolated NSs the fraction of accreting
objects is not expected to be high, and accretors are expected to be dim
(see Popov et al. 2003\nocite{agile} for a review and discussion). 

Recently \cite{kolb} discussed evolution and observational appearance
of pre-low-mass X-ray binaries (pre-LMXBs) (see also Belczynski \& Taam
2003\nocite{taam}). 
Pre-LMXBs by definition represent an evolutionary state prior to LMXB.
These systems are detached binaries in which NSs
accrete matter from the stellar wind of its companion and have 
detectable accretion luminosity.

Willems and Kolb 
found two maxima in the distribution of
accretion luminosity of NSs in pre-LMXBs
at $\sim 10^{31}$ and  $\sim10^{28}$~erg~s$^{-1}$. The second maximum
corresponds to very small accretion rates similar to those of isolated NSs
accreting the interstellar medium.
One can suspect that
for such small mass flow a NS can avoid accretion at all.
\cite{kolb} did not explore the evolutionary state of NSs in pre-LMXBs 
assuming, 
that all of them can accrete. Here in a very simple model we try to 
calculate if NSs in pre-LMXBs can reach the stage of accretion, or they are
on the propeller or ejector stages 
(see Lipunov 1992 \nocite{lip} for detailed description of different stages
and Popov et al. 2003 \nocite{agile} for a short introduction).

\section{The Model}
\label{model}

We assume that the evolution of a NS starts at the stage of ejection,
then continues to the propeller stage (including subsonic propeller,
see Davies \& Pringle 1981\nocite{davis}) 
and finally a NS can reach the stage of accretion.
Stages are separated by critical periods (see detailed derivation of them in
Lipunov 1992\nocite{lip}).
A NS is born as an ejector and starts to slow down.
The ejector stage ends when  $P=P_\mathrm{E}$:

$$
P_\mathrm{E}
=2\pi \left( \frac{2 \mu^2}{c^4V\dot M_\mathrm{acc}}\right)^{1/4}\approx
10 \, \mu_{30}^{1/2} n^{-1/4}V_6^{1/2}\, {\rm s}, R_\mathrm{l}<R_\mathrm{G}.
$$
Here $R_\mathrm{l}=cP/2\pi=4.8 \, 10^9 P\, {\rm cm}$~--~light cylinder radius,
$R_\mathrm{G}=2GM_\mathrm{NS}/V^2=3.7 \, 10^{14}V_6^{-2}\, 
{\rm cm}$~--~radius of accretion, 
$\mu_{30}=\mu/10^{30}$~G~cm$^3$~-- magnetic moment of a NS,
$V_6=V_\mathrm{Total}/10^6 {\rm cm s}^{-1}$,
$V_\mathrm{Total}=\sqrt{V_\mathrm{wind}^2+(V_\mathrm{orb})^2}$,
$n$ -- number density of the stellar wind at the NS position.
In all formulae here and below we assume NS's moment of inertia to be equal
to $10^{45}$~g~cm$^2$.

Note that the condition $R_\mathrm{l}<R_\mathrm{G}$ 
is not necessary fulfilled in the
binaries we are going to study. We discuss this topic in the Sec.~4.

Subsonic and supersonic propellers are separated by the condition of
equality of the co-rotation radius, $R_\mathrm{co}=(GMP^2/4\pi ^2)^{1/3}$,
and radius of the magnetosphere, $R_\mathrm{A}$. 
It corresponds to a critical period:

$$
P_\mathrm{ss}= 2^{5/14}\pi(GM_\mathrm{NS})^{-5/7} (\mu^2/\dot 
M_\mathrm{acc})^{3/7} \approx
$$
$$
\approx 300 \, \mu_{30}^{6/7}n^{-3/7}V_6^{9/7}\, {\rm s},\, 
R_\mathrm{A}<R_\mathrm{G}.
$$
For some systems the relation $R_\mathrm{A}<R_\mathrm{G}$
 can be violated. We discuss it below
in the Sec.~4.

Accretion starts when $P=P_\mathrm{br}$ 
(unless a NS is not on a georotator stage).

\begin{equation}
P_\mathrm{br}=
6 \,10^6\,  \mu_{30}^{16/21}\dot M_{8}^{-5/7} m^{-4/21}\, {\rm s}.
\label{pc}
\end{equation}
Here 
$\dot M_{8}=\dot M_\mathrm{acc}/10^{8}{\rm gs}^{-1}$ -- accretion rate,
$m=M_\mathrm{NS}/M_{\sun}$.
We use numerical coefficient $6\, 10^6$ 
in eq.(\ref{pc}) following Davies \& Pringle (but scaling it to
the value of $\dot M_\mathrm{acc}$ typical for pre-LMXBs).
\cite{ikh} suggested, that it should be a factor 7.5 larger.
If it is correct, then the time scale of the subsonic propeller stage
is longer. However, this stage is not well understood, and we try to be
conservative, and whenever possible choose values to lower timescales
of pre-accretion stages.

The output of our calculations consists of two timescales:
ejector time scale, $t_\mathrm{E}$, and propeller timescale, $t_\mathrm{P}$
(see Prokhorov et al. 2002\nocite{pro}, Lipunov 1992
\nocite{lip} for more details). 

At the ejector stage a NS is spinning down according to magneto-dipole
formula (approximatelly $P=3\, 10^{-4} \mu_{30} t_\mathrm{yrs}^{1/2}$). 
Propeller stage consists of two distinct phases: supersonic
propeller and subsonic propeller 
(see Davies \& Pringle  1981 \nocite{davis} and Ikhsanov 2003\nocite{ikh}),
so we define $t_\mathrm{P}=t_\mathrm{super}+t_\mathrm{sub}$.
Actually, there should be also an intermediate stage between supersonic and
subsonic propeller, when turbulence becomes supersonic
above the magnetospheric radius, $R_\mathrm{A}$, but inside 
$R_\mathrm{G}$. Spin-down 
during this intermediate phase should
be similar to the subsonic propeller rate, so they are united here. 
Supersonic propeller is a relatively short stage 
(normally shorter than each of ejector and propeller stages). 
For supersonic propeller we choose a very efficient spin-down (see Popov et
al. 2000\nocite{apj}).

Taking all together we can write the following equations for durations
of the ejector stage ($t_\mathrm{E}$), supersonic ($t_\mathrm{super}$) and
subsonic ($t_\mathrm{sub}$) propeller stages:

\begin{equation}
t_\mathrm{E}=0.8\, 10^9  \, \mu_{30}^{-1}n^{-1/2}V_6 \, {\rm yrs},
\label{te}
\end{equation}

\begin{equation}  
t_\mathrm{super}=1.3\, 10^6 \, \mu_{30}^{-8/7}n^{-3/7}V_6^{9/7}\, {\rm yrs},
\label{tsuper}
\end{equation}

\begin{equation}  
t_\mathrm{sub}=10^3 \, \mu_{30}^{-2} m\, P_\mathrm{br} \, {\rm yrs}.
\label{tsub}
\end{equation}
Eq. (\ref{te}) is taken following \cite{apj},
eq. (\ref{tsuper}) -- from \cite{shak}, and
eq. (\ref{tsub}) -- from \cite{davis}.

\begin{table}
\caption{Parameters for the models in fig.~1
\label{table}
}
\begin{tabular}{|ccccccc|}
\hline
\hline
&  & & & & &\\
$\dot M_*$, & $\dot M_\mathrm{acc}$,& $\mu_{30}$ & $T_\mathrm{orb}$, &
$V_\mathrm{wind}$, & 
$\frac{t_\mathrm{super}}{t_\mathrm{sub}}$&Mod.\\
$M_{\sun}\,{\rm yr}^{-1}$& g~s$^{-1}$&        & days       & km~s$^{-1}$ & &num.\\
\hline
& & & && & \\
$10^{-13}$ &  $3.8 \, 10^{10}$ &0.1& 1 & 660 & 0.03&1\\
$10^{-13}$ &  $3.8 \, 10^{10}$&1& 1 & 660 & 0.04&2\\
$10^{-13}$ &  $3.8 \, 10^{10}$&10& 1 & 660 &0.04 &3\\
$10^{-13}$ &  $5.2 \, 10^8$ &0.1& 30 & 660 & 0.008&4\\
$10^{-13}$ &  $5.2 \, 10^8$&1& 30 & 660 & 0.01&5\\
$10^{-13}$ &  $5.2 \, 10^8$&10& 30 & 660 & 0.01&6\\
$10^{-13}$ &  $5.0 \, 10^6$&0.1& 700 & 660 & 0.003&7\\
$10^{-13}$ &  $5.0 \, 10^6$&1& 1000 & 660 & 0.003&8\\
$10^{-13}$ &  $5.0 \, 10^6$&10& 1000 & 660 & 0.003&9\\
 & & & & & &\\
\hline
\hline
\end{tabular}
\end{table}

\begin{table}
\caption{Parameters for the models in fig.~2
\label{table}
}
\begin{tabular}{|ccccccc|}
\hline
\hline
 & & & & & &\\
$\dot M_*$, &$\dot M_\mathrm{acc}$,& $\mu_{30}$ & $T_\mathrm{orb}$, &
$V_\mathrm{wind}$, &
$\frac{t_\mathrm{super}}{t_\mathrm{sub}}$&Mod.\\
$M_{\sun}\, {\rm yr}^{-1}$& g~s$^{-1}$&  & days       & km~s$^{-1}$ & &num.\\
\hline
 & & & & &&\\
$10^{-13}$ &$3.3\, 10^{11}$ &1& 1 & 330 & 0.07&1\\
$10^{-13}$ & $3.3\, 10^{11}$&10& 1 & 330 & 0.08&2\\
$10^{-13}$ & $7.6 \, 10^9$&1& 30 & 330 & 0.02&3\\
$10^{-13}$ & $7.6 \, 10^9$&10& 30 & 330 & 0.03&4\\
$10^{-12}$ & $3.3\, 10^{12}$&1& 1 & 330 & 0.13&5\\
$10^{-12}$ & $3.3\, 10^{12}$&0.1& 1 & 330 & 0.1&6\\
$10^{-12}$ & $7.6 \, 10^{10}$&1& 30 & 330 & 0.04&7\\
$10^{-12}$ & $7.6 \, 10^{10}$&10& 30 & 330 & 0.05&8\\
$10^{-12}$ & $2.9 \, 10^{10}$&1& 1 & 1320 & 0.03&9\\
 & & & & &&\\
\hline
\hline
\end{tabular}
\end{table}

To calculate orbital velocity, $V_\mathrm{orb}$, we assume circular orbits
and equal masses $M_\mathrm{NS}=M_*=1.4\, M_{\sun}$ (Willems and Kolb 
\nocite{kolb}
discuss $M_{\sun}<M_*<2\, M_{\sun}$). In our calculations the
orbital velocity is always smaller than the stellar wind velocity.

To calculate the accretion rate we use the following equation (see
Willems \& Kolb 2003\nocite{kolb}):

$$
\dot M_\mathrm{acc}=
\frac{3}{16}\left(\frac{R_*}{a}\right)^2 \frac{q^2}{\beta^4}
\left( 1+ \frac{1+q}{2 \beta^2} \frac{R_*}{a}\right)^{-3/2} \dot M_*.
$$
Here $R_*$ -- stellar radius,
$q=M_\mathrm{NS}/M_*$, $\dot M_*$ -- the rate of stellar wind mass loss. 
We used several different values for $\dot M_*$ and 
$V_\mathrm{wind}=\beta \, \sqrt{2GM_*/R_*}$ (see tables,
330 km~s$^{-1}$ corresponds to $\beta=0.5$, 660  km~s$^{-1}$ -- to
$\beta=1$ and 1320  km~s$^{-1}$ -- to $\beta=2$). 
Note, that actual accretion rate can be lower than this value 
due to heating, magnetospheric and hydrodynamical effects
(see discussion in Perna et al. 2003\nocite{perna} and Popov et al.
2003\nocite{agile}).

To derive many equations in this paper we will use the following relation
which parametrises the accretion rate by $n$ and $V_\mathrm{Total}$:

$$
\dot M_\mathrm{acc}=7\, 10^{11}\, n\, V_6^{-3}\, {\rm g s}^{-1}.
$$
$n=\rho m_\mathrm{p}^{-1}$,
$\rho=\dot M_*/(4\pi a^2 V_\mathrm{wind})$ -- density of matter, 
$m_\mathrm{p}$ -- proton mass.

For $\dot M_*$  on the main sequence stage \cite{kolb} used the value
$10^{-13}$~$M_{\sun}$~yr$^{-1}$. We follow them in that choice 
and also check higher value $10^{-12}$~$M_{\sun}$~yr$^{-1}$, which was also
discussed by this authors.
Definitely for smaller $\dot M_*$ all timescales ($t_\mathrm{E}$, 
$t_\mathrm{super}$,
$t_\mathrm{sub}$) become longer, for higher $\dot M_*$ -- shorter.

Mass loss can be increased significantly when a donor star leaves the main
sequence, but we do not follow these stages in our analysis.
Also we do not discuss the possibility of a formation of a binary system due
to capturing of a companion (in that case the picture can be completely
different, as far as an old evolved NS can be captured).

We want to note once again that our model is a very simple one as far as we
want just to illustrate the importance of the effects of magnetorotational
evolution of NSs in pre-LMXBs. Discussion on possible complifications can be
found below in the Sec.~4.

\section{Results}
\label{results}

Our main results are presented in the figures. 
Each point represents a model with particular set of parameters
(see  tables). The three lines correspond to the total time
of evolution prior to the stage of accretion 
($t_\mathrm{Total}=t_\mathrm{E}+t_\mathrm{P}$)
equal to 1, 5 and 10 billion years. 

Note, that $t_\mathrm{P}\propto t_\mathrm{E}^a$, $a\approx 1.3$.  
This happens because of the following reason.
In eq.~(\ref{te}) $t_\mathrm{E}\propto \mu^{-1} n^{-1/2}$ and 
$V_\mathrm{Total}$ is not varied
significantly in our modeling as far as $V_\mathrm{wind}>V_\mathrm{orb}$.
If we substitute eq.~(\ref{pc}) into eq.~(\ref{tsub}), then we have
$t_\mathrm{sub}\propto \mu^{-26/21} n^{-5/7}$. (Remember, that in
$t_\mathrm{P}$
we have $t_\mathrm{super} << t_\mathrm{sub}$ [see tables].) As we see we have nearly
$t_\mathrm{P}\propto t_\mathrm{E}^{1.3}$.

Vertical line in the figures corresponds to $t_\mathrm{E}=2.3 \, 10^9$~yrs
 -- the main sequence lifetime of a star with
$M_*=1.4\, M_{\sun}$. (Roughly, the main sequence lifetime can be estimated
as: $\log t_\mathrm{MS}=9.9-3.8\log(M_*/M_{\sun})+\log^2(M_*/M_{\sun})$,
see for example Lipunov 1992\nocite{lip}.)
If for some reason the propeller stage appears to be very short
(for example, subsonic propeller is not operating) then NSs to the left of
this line can accrete.

As one can see most of models from table~1 have 
$t_\mathrm{Total}>2.3 \, 10^{9}$~yrs.
It means that these NSs can not accrete. Accretion is more probable onto
strongly magnetized NSs as far as their spin-down timescales are shorter
on both ejector and propeller stages. 
To allow accretion with a rate $\sim10^{10}$~g~s$^{-1}$
a NS with $B\sim 10^{12}$~G should be slowed down to very long periods
$\sim 10^5$~s which takes a long time.

On the second figure we present models with $t_\mathrm{Total}<5\, 10^{9}$~yrs.
These are systems with strongly magnetized NSs or/and with shorter orbital
periods or/and with smaller wind velocity.

\section{Discussion and conclusions}
\label{discuss}

\begin{figure}
\vbox{\psfig{figure=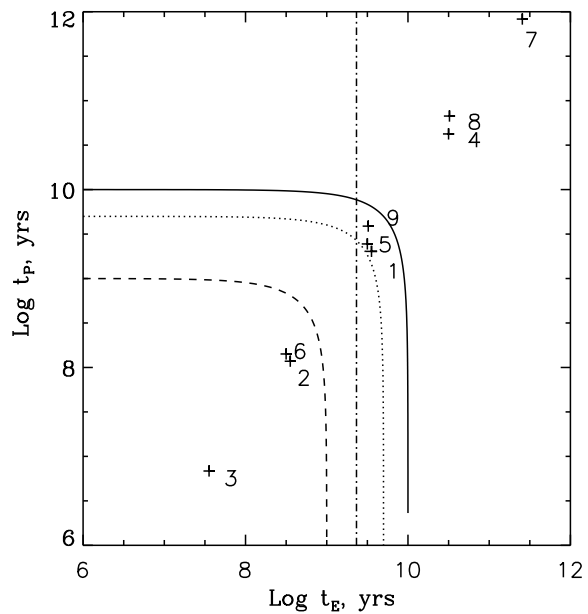,width=\hsize}}
\caption[]{ Crosses with numbers represent different set of parameters of NSs
(see table~1).
Three curves corresponds to $t_\mathrm{Total}=t_\mathrm{E}+t_\mathrm{P}$ 
equal to 1 Gyr (dashed),
5 Gyrs (dotted) and 10 Gyrs (solid).The vertical line corresponds to
$t_\mathrm{E}=2.3 \, 10^9$~yrs.
}
\label{fig:pre}
\end{figure}

\begin{figure}
\vbox{\psfig{figure=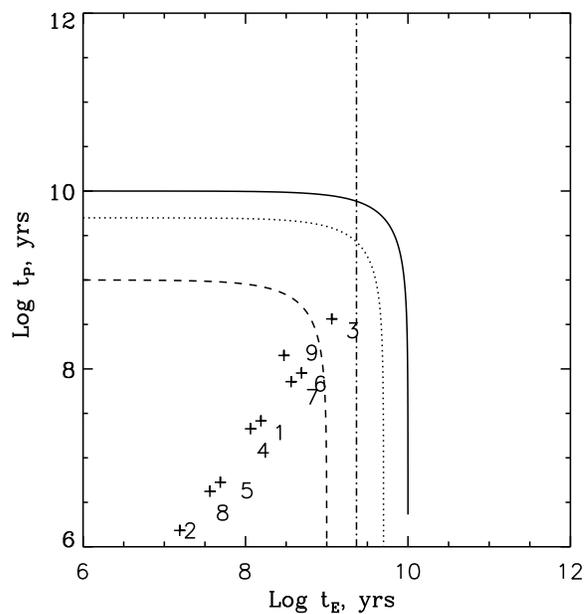,width=\hsize}}
\caption[]{ Same as in fig.~1 for parameters in table~2.
}
\label{fig:pre}
\end{figure}

Our calculations were made for constant magnetic fields.
As it was shown by \cite{colpi}, \cite{livio}, \cite{pp}
realistic parameters of decay in the case of isolated NSs
normally make the propeller stage longer. 
Parameters of the magneto-rotational evolution of isolated NSs
and NSs in pre-LMXBs are very similar (for example, there is no huge
accretion to speed-up the process of field decay), 
so the results for the former ones can be applied to the latter
(and vice versa).
In that sense our estimates of $t_\mathrm{E}$ 
and $t_\mathrm{P}$ for $B=const$
should be lower limits if the field is decaying.
For decayed fields most of NSs in pre-LMXBs can be expected to stay on the
propeller stage.
For high $V_\mathrm{wind}$ a NS instead of accretion
can stay on the so-called georotator stage (Lipunov 1992\nocite{lip}).
It happens for $V_\mathrm{Total}\ga 470 \, \mu_{30}^{-1/5} n^{1/10}\, 
{\rm km s}^{-1}$ (see Popov et al. 2003\nocite{agile}).

In this short illustrative note we neglect several effects, which can
change characteristic periods and radii or/and spin--down rates.
Most of the time scales used above can be considered just as rough estimates
(see discussion in Popov et al. 2003\nocite{agile}).
However, we believe that qualitatively our results are valid as far as our
main aim is only 
to illustrate importance of the effects of magneto-rotational evolution of
NSs in pre-LMXBs.

Let us briefly discuss possible violations of the relations
$R_\mathrm{l}<R_\mathrm{G}, \, R_\mathrm{A}<R_\mathrm{G}$ 
(for details see Lipunov 1992\nocite{lip}). 
We start with the relation between $R_\mathrm{l}$ and $R_\mathrm{G}$.
For high magnetic fields $P_\mathrm{E}$ 
becomes large. It necesserely means large
$R_\mathrm{l}$. On the other hand $V_\mathrm{wind}$ 
is also large, which means small $R_\mathrm{G}$.
For systems 3,  6,  8, 9 in the table~1 and for system  9 in the
table~2 $R_\mathrm{l}>R_\mathrm{G}$ at the moment of transition from ejector
to propeller. 
In that case equations for $P_\mathrm{E}$ and $t_\mathrm{E}$ are 
the following:

\begin{equation}
P_\mathrm{E2}\approx 180\, \mu_{30}^{1/3} n^{-1/3} V_6^{-1/3}{\rm s},
\label{pe2}
\end{equation}

\begin{equation}
t_\mathrm{E2}\approx 3.5 \, 10^{11} \mu_{30}^{-4/3} n^{-1/3} V_6^{-2/3}\, {\rm yrs}.
\label{te2}
\end{equation}

For large velocities and magnetic fields normally 
$t_\mathrm{E}>t_\mathrm{E2}$, so the stage
of ejection is shorter for $R_\mathrm{l}>R_\mathrm{G}$
(compare eqs.~\ref{te}, \ref{te2}). 
For our investigation it is crucial to know if this
effect is strong enough to let the system to cross the line $t_\mathrm{E}=2.3 \,
10^9$~yrs on the figures. Only for the model 9 from the table~1 it is so
(in the figures all data point are plotted for $R_\mathrm{l}<R_\mathrm{G}$).
For all others this shortening of $t_\mathrm{E}$ is not strong enough to change
the conclusions made in this short note. Also we have to mention that 
if a system leaves the ejector stage after the violation 
$R_\mathrm{l}<R_\mathrm{G}$ then the
accretion rate is expected to be very low.

Now let us discuss the relation between $R_\mathrm{A}$ and $R_\mathrm{G}$.
Often $R_\mathrm{A}<R_\mathrm{G}$ 
(for example it is so for all observed bright X-ray
binaries, for isolated NSs which can become accretors etc.).
For systems with highly magnetized NSs and large wind velocities it is
possible, that $R_\mathrm{A}>R_\mathrm{G}$. It is important for example to
estimate the critical period $P_\mathrm{ss}$ and $t_\mathrm{sub}, 
\, t_\mathrm{super}$. 
Alfven radius is determined by the following equations:

$$
R_\mathrm{A} = 
\left( \frac{\mu^2}{2\dot M_\mathrm{acc}\sqrt{GM_\mathrm{NS}}}\right)^{2/7}\approx
$$

$$
8.2 \, 10^9 \mu_{30}^{4/7} n^{-2/7} V_6^{6/7} \, {\rm cm},\, 
R_\mathrm{A}<R_\mathrm{G},
$$

$$
R_\mathrm{A2}= 
\left( \frac{4 \mu^2 G^2M_\mathrm{NS}^2 }{\dot M_\mathrm{acc} V^5}\right)^{1/6}\approx
$$

$$
7.6 \, 10^{11} \mu_{30}^{1/3} n^{-1/6} V_6^{-1/3}\, {\rm cm},\, 
R_\mathrm{A2}>R_\mathrm{G}.
$$
In the later case the equation for the critical period is:

$$
P_\mathrm{ss2}= 
2\pi \left( \frac{2\mu^2}{\dot M_\mathrm{acc} V^5}\right)^{1/4}\approx
2.5 \, 10^6 \mu_{30}^{1/2} n^{-1/4} V_6^{-1/2} \, {\rm s}.
$$
  
For systems with $R_\mathrm{A}>R_\mathrm{G}$ 
durations of the propeller substages are different
from eqs.(\ref{tsuper}, \ref{tsub}). These systems later  appear 
not as normal accretors but as georotators 
and so we do not take this effect into account.  
Study of georotators in binaries is a
separate (and probably promissing) subject. At that stage there is a
possibility of a particular "magnetic accretion", see \cite{rut}.

In the discussed type of binaries there are possibilities for episodes of
accretion.
As far as low-mass stars can have episodes of eruptive activity  it is 
possible that NSs can accrete matter due to fluctuations in the stellar
wind. 
Post main sequence stages are relatively short, 
so no significant spin-down is expected. 
There is just a quick increase of the accretion rate: 
increase of $\dot M_*$ and decrease of $V_\mathrm{wind}$ both work to
increase $\dot M_\mathrm{acc}$.  
Non-accreting systems, which are situated in $t_\mathrm{E}$-$t_\mathrm{P}$-plane
close to the line $t_\mathrm{E}=2.3 \, 10^9$~yrs 
(like model 1,5,9 in the fig.~1)  can start to accrete
if $\dot M_\mathrm{acc}$ is increased when the star leaves the main sequence but the
system is still detached.
Systems with highly eccentric orbits can have episodes of accretion in
periastron.

We conclude, that as in the case of isolated NSs many  NSs in pre-LMXBs
(at least on the main sequence)
are not at the stage of stable wind accretion.
Accretion is possible when the donor is  on the stage of main sequence
if a NS is strongly magnetized or in close binaries. 

\begin{acknowledgements}
The author thanks Roberto Turolla and Aldo Treves for discussions and 
detailed comments on the text and the unknown referee for helpful criticism.
\end{acknowledgements}

\end{document}